\magnification=\magstep2
\def\para{\par\noindent}
\def\sqr#1#2{{\vcenter{\vbox{\hrule height.#2pt
        \hbox{\vrule width.#2pt height#1pt \kern#1pt
          \vrule width.#2pt}
        \hrule height.#2pt}}}}

\newcount\notenumber

\def\note{\advance\notenumber by 1
\footnote{$^{\the\notenumber}$}}
\baselineskip 20pt
\centerline{{\bf The Four-Dimensional XY Spin Glass}}\para
\vskip 0.25cm
 \para S.~Jain,
 \para School of Mathematics and Computing,\para
 University of Derby,\para
 Kedleston Road,\para
 Derby DE22 1GB,\para
 U.K.\para
\vskip 0.25cm
\para E-mail: S.Jain@derby.ac.uk
\vskip 2.0cm
\para Classification Numbers: 0570J, 6460C, 7510H, 7510N
\vfill\eject
\para ABSTRACT
\par The nearest-neighbour XY spin glass on a hypercubic lattice in
 four dimensions is studied by
Monte Carlo simulations. A finite-size scaling analysis of the data leads to
 a finite temperature spin glass
transition at $T_c=0.95\pm 0.15$. The critical exponents are estimated to be
$\nu_{sg}=0.70\pm 0.10$ and $\eta_{sg}=-0.28\pm 0.38$.
 The results imply that the lower
 critical dimensionality for the XY spin glass is less than four.
\vfill\eject
\para 
Recently there has been considerable interest in the behaviour of short-range
 vector spin glasses [1-8]. Although there is now convincing evidence
 [2-4] that
XY spin glasses exhibit only a zero-temperature phase transition for $d=2$
and 3, the location of the lower critical dimension, $d_L$, remains 
controversial.
A zero-temperature study by Morris et al [2] suggests that $d_L=4$. The
validity of Nishimori
and Ozeki's [5] attempt at a Mermin [9] type argument has been
 questioned [6] and it is claimed by
 Schwartz and Young [7] that all one can actually show is
 that $d_L\ge 2$.

A recent Migdal-Kadanoff renormalisation-group analysis [8], on the other hand,
claims that the XY spin glass orders at a finite temperature for $d=4$ and,
therefore, $d_L < 4$.
The XY spin glass, unlike the corresponding gauge glass,
 possesses a \lq reflection\rq\ 
symmetry.
It has therefore been argued [10, 11] that the two models belong to different
universality classes. Recent computer simulations [11] of the gauge glass
 in four
dimensions clearly show a finite temperature transition.

As far as we are aware, to-date the XY spin glass has not been studied
by simulations for $d=4$. We attempt to fill this gap by presenting
in this letter the first results of extensive Monte Carlo simulations
of the four dimensional XY spin glass with $\pm J$ nearest neighbour
interactions. 
Applying finite-size scaling theory [12], we shall find evidence for a 
finite-temperature glass transition.
\para
The Hamiltonian for the model simulated is given by
$$ {\it H} =-\sum_{<i,j>} J_{ij}{\bf S}_i.{\bf S}_j=-\sum_{<i,j>} J_{ij} \cos(
\theta_i - \theta_j),\eqno(1)$$
with $0\le\theta_i\le 2\pi$ for all $i$. The planar spins,
 ${\bf S}_i (=(S_{i,x},S_{i,y}))$,
 are situated on
every site of a four dimensional hypercubic lattice of size $L^4 (L=2,4$ and
 $6)$.
 The
 summation runs over nearest-neighbour pairs only and the
interactions, $J_{ij}$, are independent random variables selected from a
binary $\pm 1$ distribution. As usual, the temperature is given in
 units of the nearest neighbour interaction.
We work with full periodic boundary conditions.

In the simulations we study the spin-glass susceptibility, $\chi_{sg}$, which 
is defined by
$$\eqalignno{\chi_{sg}& = {1\over N}\sum_{i,j}[<{\bf S}_i.{\bf S}_j>^2_T]_J\cr
     &=N q^{(2)}_{sg},&(2)\cr}$$
where $<...>_T$ indicates a thermal average, $[...]_J$ denotes an average
over the disorder, $N=L^4$ and
$$q^{(2)}_{sg}=\sum_{\mu ,\nu}[<q^2_{\mu\nu}>_T]_J.\eqno(3)$$
In equation (3) the tensor variable, $q_{\mu\nu}$, is defined in terms of the
overlap between two replicas 1 and 2,
$$q_{\mu\nu} = {1\over N}\sum_{i}S^{1}_{i,\mu}S^{2}_{i,\nu}\qquad 
(\mu ,\nu = x,y).\eqno(4)$$
Higher order correlations such as $q^{(4)}_{SG}$ can also be written
 in terms of $q_{\mu\nu}$, namely
$$ q^{(4)}_{sg}=\sum_{\mu ,\nu ,\alpha ,\beta}[<q^2_{\mu\nu} 
q^2_{\alpha\beta}>_T]_J.\eqno(5)$$
Rather than use $q^{(2)}_{sg}$ and $q^{(4)}_{sg}$, it is far more
 convenient to work with the dimensionless Binder parameter [12] defined
by
$$ g_{sg} = 3 - 2{{q^{(4)}_{sg}}\over {(q^{(2)}_{sg})^2}}.\eqno(6)$$
According to finite-size scaling theory [12], near $T_c$ we expect the
 Binder parameter
 to scale as
$$g_{sg}(L,T)=\overline g_{sg}(L^{1/\nu_{sg}}(T-T_c)), \eqno(7)$$
where $\nu_{sg} $ is the correlation length exponent and $\overline g_{sg}$
is a scaling function satisfying
$$\overline g_{sg}(L^{1/\nu_{sg}}(T-T_c))=\cases{0&\qquad for $T>T_c, L\rightarrow\infty$\cr
                          1&\qquad for $T<T_c, L\rightarrow\infty$\cr}\eqno(8)$$
provided that the ground state is non-degenerate. Whereas
 for a finite-temperature
spin glass transition plots of $g_{sg}(L,T)$ versus $T$ for different $L$
should intersect at $T_c$, for a transition at zero temperature we expect the
curves to meet each other at $T=0$ only.

The scaling form for the spin glass
 susceptibility is given by
$$ \chi_{sg}(L,T)=L^{2-\eta_{sg}}\overline \chi_{sg}(L^{1/\nu_{sg}}(T-T_c)), 
\eqno(9)$$
 where the exponent $\eta_{sg} $ describes the power-law decay of correlations
at the transition temperature and $\overline \chi_{sg}$ is now a scaling
function. It follows from equation (9) that
$$\chi_{sg}(L,T_c)\sim L^{2-\eta_{sg}}.\eqno(10)$$
As the co-ordination number for our model is $Z=8$, the mean-field
 values of $T_c$ and the 
exponents mentioned above are 
$$ T^{mf}_c\approx\sqrt Z/2=\sqrt 2, \qquad\nu^{mf}=1/2\qquad\hbox{and}
\qquad \eta^{mf}=0.\eqno(11)$$  
We now describe our Monte Carlo simulations and discuss the results.
During the simulations, which were performed using the conventional Metropolis
[13] technique on a Cray Y-MP8 and a J90, we actually work with discrete spins.
For technical reasons, the spins were discretized to occupy 256 equally spaced
orientations in the plane. Furthermore, a variant of multispin coding [14] was 
used to store 7 (discrete) spins in one word and the lattice was composed of
two inter-penetrating sub-lattices. As a consequence, each update of the
 lattice allows us to update 14 separate samples (or, alternatively, 7 pairs of
samples) at the same time.

We follow Bhatt and Young [12] and compare the spin glass correlations
obtained from 2 independent replicas with the same set of bonds with those from
a single replica at 2 different times. Equilibrium is assumed only if the values
agree within the statistical error.

The number of Monte Carlo steps, $\tau_0(L,T)$, required to achieve equilibrium
depends on both the system size $L$ and the temperature $T$. $\tau_0(L,T)$ sets
 upper and lower limits on the values of $L$ and $T$, respectively, that
 can be studied. In our simulations we found that $\tau_0(2,0.4)\approx 1000, 
\tau_0(4,0.7)\approx 5000$ and $\tau_0(6,0.75)\approx 13000$ sweeps;
 equilibration
 problems
prevented us from going to lower temeperatures.
The number of bond configurations we generated to average
over the disorder also varied with $L$. Typically, we considered $7000 (L=2),
250\sim 2300 (L=4), 100\sim 500 (L=6)$ pairs of samples for each
 temperature considered.
(However, note that for the lowest 2 temperatures ($T=0.775$ and $0.75$) 
for $L=6$ we averaged over only 56 pairs of samples in each case.)  
In total, the simulations presented in this work took approximately 600 hours
of CPU time on the two supercomputers mentioned above.

In figure 1 we plot $\chi_{sg}$ against $T$ for the 3 different values of $L$
considered. The statistical error-bars were evaluated from the sample-to-sample
fluctuations and are only displayed in those cases where they exceed the size
of the points.

The Binder parameter is plotted against the temperature in figure 2. Although
the curves clearly intersect at a finite temperature, the point of intersection
is not unique. This is probably due to corrections to finite-size scaling and
the statistical error in $g_{sg}$. As we have data for 3 values of $L$,
 we obtain 3 intersection
temperatures, $T^{L_1,L_2}_c$, where $L_1, L_2 = 2, 4, 6$ and $L_1\ne L_2$.
 For the data presented in figure 2 we note 
that there is a small downward shift in the value of $T^{L_1,L_2}_c$ for 
increasing $L_1$ and $L_2$. Clearly, to establish whether the shift is
 significant or not, it is highly desirable to obtain additional
data for the Binder parameter for larger lattices and lower temperatures.
 However, we note that there is
some evidence of a finite-temperature transition as the curves clearly splay out
 below the intersection point. From figure 2 we
 estimate the spin-glass transition temperature to be $T_c=0.95\pm 0.15$. Our
value of $T_c (\approx 0.7\ T_c^{mf})$ agrees well
 with the value of $T_c\approx 0.9$ obtained recently by
Nobre et al [8]. It is also surprisingly close to the transition temperature
found by Reger and Young [11] for the four dimensional {\it gauge} glass.

Having obtained an estimate for $T_c$, a log-log plot of
 $\chi_{sg}(L,T_c)$
against $L$ is expected from 
equation (10) to have a slope of $2-\eta_{sg}$. Our results are 
consistent
with this but the uncertainty in $T_c$ leads to a large error in $\eta_{sg}$ and we
estimate that $2-\eta_{sg}=2.28\pm 0.38$.

To fix the second independent exponent, $\nu_{sg}$, we display in figure 3
a scaling plot of $\chi_{sg}(L,T)/L^{2.28}$ against $(L^{1/\nu_{sg}}(T-0.95))$.
 To
see how sensitive the scaling plot is to the value of $\nu_{sg}$, we have tried 
a range of values. As a result we estimate $\nu_{sg}=0.70\pm 0.10$. 
As can be seen from figure 3, the data scale reasonably well for $\nu_{sg}=0.70$.
 Once again, our value for the correlation
 exponent is remarkably close to the one found for the gauge glass in 4d [11]. 

Finally, in
figure 4 we show a scaling plot of the data for the Binder parameter for the
{\it same} values of $T_c$ and $\nu_{sg}$ as the ones used in the plot for
figure 3. We see that the data for $g_{sg}$ do not scale as well as those for
$\chi_{sg}$. The quality of the data collapse does not improve for other possible
values of $T_c$ and $\nu_{sg}$. (A correction to scaling as suggested by Bokil
and Young [15] also fails to make any significant difference to the plot.) A
 similar behaviour was found by Kawamura [4] in
the three-dimensional XY spin glass.  

It has been assumed that the XY spin glass and the gauge glass belong to
different universality classes as the latter does not share the reflection
symmetry of the former. As noted, our results are remarkable in their similarity
to those found earlier by Reger and Young [11] for the vortex glass in four
 dimensions. It is felt that this unexpected feature requires further
investigation.

To conclude, we have presented numerical evidence that the XY spin glass has
a finite temperature glass transition in four dimensions. We have estimated the
critical temperature and the critical exponents. Further work is required to
confirm the transition temperature and obtain more accurate values for
 the exponents. Our results are in
agreement with the analytic approximation carried out by Nobre et al [8].
They are also surprisingly similar to those found earlier
 by Reger and Young [11] for the
four dimensional gauge glass and imply that $d_L < 4$ for the XY
spin glass.

Work is underway to investigate the chiral-glass [4, 15] behaviour
 of the model using
the vortex representation [15].

The simulations were performed on a Cray YMP and a J90
 at the Rutherford Appleton 
Laboratory through an Engineering and Physical Sciences Research Council 
(EPSRC) research grant (Ref: GR/K/00813). 
\vfill\eject 
\para FIGURE CAPTIONS
\vskip 1cm
\para Figure 1

\para A plot of the spin glass susceptibility, $\chi_{sg}$, against the
temperature for $L=2, 4$ and $6$. The lines are just guides to the eye.
\vskip 1cm

\para Figure 2

\para A plot of the Binder parameter defined in equation (6) against the
 temperature for $L=2, 4$ and $6$. The lines are just
 guides to the eye.
\vskip 1cm

\para Figure 3

\para A scaling plot of $\chi_{sg}/L^{2-\eta_{sg}}$ versus $L^{1/\nu_{sg}}(T-T_c)$
with $\eta_{sg}=-0.28, \nu_{sg}=0.70$ and $T_c=0.95$. See equation (9) in the 
text.
\vskip 1cm

\para Figure 4

\para A scaling plot of the Binder parameter $g_{sg}$ versus
 $L^{1/\nu_{sg}}(T-T_c)$
for $\nu_{sg}=0.70$ and $T_c=0.95$.
\vfill\eject
\para REFERENCES
\item {[1]} Banavar J R and Cieplak M 1982 Phys. Rev. Lett. {\bf 48}
832
\item {} McMillan W L 1985 Phys. Rev. B {\bf 31} 342
\item {} Olive J A, Young A P and Sherrington D 1986 Phys. Rev. B {\bf 34}
 6341
\item {} Matsubara F and Iyota and Inawashiro 1991 Phys. Rev. Lett. {\bf 67}
1458
\item {} Kawamura H 1992 Phys. Rev. Lett. {\bf 68} 3785; 1995 J. Phys. Soc. Jpn.
{\bf 64} 26
\item {} Coluzzi B 1995 J. Phys. A: Math. and Gen. {\bf 28} 747
\item {[2]} Morris B W, Colborne S G, Moore M A, Bray A J and Canisius 1986
J. Phys. C: Solid State Phys. {\bf 19} 1157 
\item {[3]} Jain S and Young A P 1986 J. Phys. C: Solid State Phys. {\bf 19}
 3913
\item {} Ray P and Moore M A 1992 Phys. Rev. B {\bf 45} 5361
\item {[4]} Kawamura H 1995 Phys. Rev. B {\bf 51} 12398
\item {} Kawamura H and Tanemura M 1985 J. Phys. Soc. Jpn. {\bf 54} 4479; 1986
  J.Phys. Soc. Jpn. {\bf 55} 1802
\item {[5]} Nishimori H and Ozeki Y 1990 J. Phys. Soc. Jpn. {\bf 59} 295
\item {} Ozeki Y and Nishimori H 1992 Phys. Rev. B {\bf 46} 2879
\item {[6]} O'Neill J A and Moore M A 1990 J. Phys. Soc. Jpn. {\bf 59} 289
\item {[7]} Schwartz M and Young A P 1991 Europhys. Lett. {\bf 15} 209
\item {[8]} Nobre F D, Mariz A M and Sousa E S 1993 Physica A {\bf 196} 505
\item {[9]} Mermin N D 1967 J. Math. Phys. {\bf 8} 1061
\item {[10]} Reger J D, Tokuyasu T A, Young A P and Fisher M P A 1991 Phys. Rev.
 {\bf 44} 7147
\item {} Fisher M P A, Tokuyasu T A and Young A P 1991 Phys. Rev. Lett.
 {\bf 66} 2931
\item {} Huse D A and Seung H S 1990 Phys. Rev. B {\bf 42} 1059
\item {[11]} Reger J D and Young A P 1993 J. Phys. A: Math. and Gen. {\bf 26}
 L1067
\item {[12]} Bhatt R N and Young A P 1988 Phys. Rev. B {\bf 37} 5606
\item {[13]} Metropolis N, Rosenbluth A W, Rosenbluth M N, Teller A H and
 Teller E 1953 J. Chem. Phys. {\bf 21} 1087
\item {[14]} Bhanot E, Duke D and Salvador R 1986 J. Stat. Phys. {\bf 44} 85;
 1988 Comput. Phys. Commun. {\bf 49} 465
\item {[15]} Bokil H S and Young A P 1996 J. Phys. A: Math. and Gen. {\bf 29}
 L89
\end